\def\breakon{\end{multicols}\widetext\vspace{-.60cm}
\noindent\rule{.49\linewidth}{.3mm}\rule{.3mm}{.5cm}\vspace{0.0cm}}
\def\breakoff{\vspace{-0.5cm}
\noindent
\vspace{0.0cm}
\begin{multicols}{2}
\narrowtext}
\documentstyle[aps,epsf,prl,multicol]{revtex}
\begin{document}

\title{ Spin symmetry breaking in bilayer quantum Hall systems}
\author{ Eugene Demler$^{1}$, Eugene H. Kim$^{2}$, and S. Das Sarma$^{3}$}
\address{$^1$ Institute for Theoretical Physics, University of
California, Santa Barbara, CA 93106}
\address{$^2$ Department of Physics, University of
California, Santa Barbara, CA 93106}
\address{$^3$ Department of Physics, University of Maryland, College
Park, MD 20742}
\maketitle

\begin{abstract}
Based on the construction of generalized Halperin wave functions, we
predict the possible existence of a large class of broken spin
symmetry states in bilayer quantum Hall structures, generalizing
the recently suggested canted antiferromgnetic phase to many
fractional fillings. We develop the appropriate Chern-Simons
theory, and establish explicitly that the low-lying neutral
excitation is a Goldstone mode and that the charged excitations are
bimerons with continuously tunable (through the canted
antiferromagnetic order parameter) electric charge on the individual
merons.


\end{abstract}

\vspace{-.4in}
\begin{multicols}{2}

Recently, a canted antiferromagnetic (CAF) state has been
predicted to exist in bilayer quantum Hall (QH) systems at
the special filling factor $\nu=2$ \cite{R1}, or more generally at
$\nu=2/m$ where $m$ is an odd integer \cite{R2}. The original theoretical
prediction \cite{R1} based on a microscopic Hartree-Fock calculation
has been followed up by a number of subsequent theoretical works using
a quantum nonlinear sigma model \cite{R2}, a bosonic spin approach
\cite{R3,R3a}, and more detailed Hartree-Fock calculations
\cite{R4,R5,R6}. Fairly persuasive experimental support for
the CAF phase in $\nu=2$ bilayer QH systems also exists \cite{R7}. The basic
idea underlying the novel CAF phase is that the
competition between interlayer tunneling, Zeeman splitting,
intralayer Coulomb interactions, and interlayer Coulomb interactions can
cause spontaneous symmetry breaking in bilayer systems, leading to the CAF
phase.  This CAF phase lies in between the usual
spin polarized ferromagnetic phase and the symmetric paramagnetic
(or singlet) phase.

In spite of the extensive theoretical work on the problem using
Hartree-Fock or related spin operator approaches
\cite{R1,R2,R3,R3a,R4,R5,R6}, a
fundamental understanding of the precise nature of the CAF phase,
either from the perspective of actual QH wave functions or from a
long wavelength field theoretic viewpoint, is still lacking.
In this Letter we construct a
microscopic wave function for the ground state of the CAF state,
and develop a Chern-Simons theory to study the excitations above the
CAF ground state.  We find a neutral Goldstone mode associated with the
breaking of the spin symmetry in the CAF phase and novel bimeronic charged
excitations, which we discuss below.  Furthermore, we establish that
the type of symmetry breaking characterizing the CAF state is quite
generally allowed in bilayer QH systems and may in principle exist for a
large class of QH states far beyond the originally predicted $\nu=2/m$
filling factors.

As pointed out first by Wen and Zee \cite{WenZee}, the Halperin
$(m,m,m)$ wave functions \cite{Halperin} (neglecting electron spin)
in bilayer QH systems have the property of fixing the total filling
factor $\nu$ in the system, but not the individual filling factors
of each layer.  This allows one to construct wave functions that
are a superposition of states with different numbers of particles
in each layer (but fixed total number of particles), which leads to
spontaneous interlayer coherence in the absence of interlayer tunneling
\cite{WenZee,Indiana1}.  We show that the analogous Halperin-type
construction for spinful electrons in bilayer QH systems leads to a CAF
phase that breaks the spin symmetry spontaneously.

The electron Hamiltonian for a bilayer QH system can be written as
\begin{eqnarray}
{\cal H} &=& \int d^2 x \left\{ \frac{1}{2 m } \left | (
 -i \vec{\partial} - e {\bf A}^{ex} ) \Psi_{a \sigma} \right|^2
 \right. \nonumber \\
 & + & \left. (u^c_I + u^c_O ) \left( \bar{\Psi}_{a\sigma} \Psi_{a\sigma}
\right)^2 + \Delta_{Z} \bar{\Psi}_{a\alpha} \sigma^z_{\alpha\beta}
\Psi_{a\beta}  \right. \nonumber\\
 &+& \left. (u^c_I - u^c_O ) \left(
\bar{\Psi}_{a\sigma} \tau^z_{ab} \Psi_{b\sigma} \right)^2 +
\Delta_{SAS} \bar{\Psi}_{a\sigma} \tau^x_{ab} \Psi_{b\sigma} \right\} \, .
\end{eqnarray}
Here $a$, $b$ and $\alpha$, $\beta$ are layer (''isospin'') and spin
indices, respectively; $u_I^c$ is the intralayer and $u_O^c$ is the
interlayer Coulomb interaction; $\Delta_{SAS}$ is the splitting between
symmetric and antisymmetric states; $\Delta_z$ is the Zeeman splitting.
In the case when $u^c_I = u^c_O $ (i.e. $d=0$, where $d$ is the interlayer
distance) and $\Delta_{SAS}=0$, the system has a full isospin $SU(2)$ symmetry
generated by $ T^{q} = \int d^2x \bar{\Psi}_{a\sigma} \tau^q_{ab}
\Psi_{b\sigma} $. When only one of $d$ or $\Delta_{SAS}$ is finite, the
isospin symmetry reduces to a smaller $U(1)$ symmetry generated by
either $T^z$ or $T^x$. It is then reasonable to define single particle
wave functions having certain transformation properties under the
explicit symmetry of the Hamiltonian.  For $T^z$ symmetry such
wave functions would describe electrons in a specific layer; for $T^x$
they are symmetric and antisymmetric combinations of electrons in the
two layers.  In general (and in the relevant experimental systems)
both $d$ and $\Delta_{SAS}$ are finite.  Therefore, any $U(1)$ symmetry is
destroyed, and there is no natural choice of single electron
wave functions. We choose to work with the symmetric/antisymmetric
electron wave functions, since in the absence of a gate voltage between
the layers, $\bar{\Psi}_{a\sigma} \tau^z_{ab} \Psi_{b\sigma}$ does not
acquire an expectation value and has smaller effect on single electron
states than interlayer tunneling.
Our approach is well justified as long as $\Delta_{SAS}$ is
not much smaller than $u^c_I - u^c_O$. In the
presence of interlayer tunneling, there is no degeneracy between
different single electron states. Of the four possible single particle
states, the symmetric spin up ($S\uparrow$) state always has the lowest
energy and antisymmetric spin down ($A\downarrow$) has the highest energy.
Since the symmetric spin down ($S\downarrow$) and the antisymmetric
spin up ($A\uparrow$) states may be close in energy, Coulomb
interactions may lead to considerable mixing between them.

We now construct a Halperin-like wave function \cite{Halperin}
for our spinful bilayer system.  Our wave function does not fix
the number of electrons in the $S\downarrow$ and $A\uparrow$
states individually, but fixes their sum.  Let us first assume
that the highest energy $A\downarrow$ state is completely
empty \cite{note}. Then, if we label $S\uparrow$ states by $z$,
$S\downarrow$ states by $u$, and $A\uparrow$ states
by $w$, we can easily write the Halperin wave function
that fixes the number of electrons in the $S\downarrow$ and $A\uparrow$
states together, but not in each of them separately:
\begin{eqnarray}
 & & \Psi(\{z\} \{u\} \{w\}) = \prod ( z_i - z_j)^n  \prod ( z_i - w_j)^l
 \prod (z_i - u_j)^l
 \nonumber \\
 & & \hspace{.35in} \times \prod (w_i - w_j)^m \prod (w_i - u_j)^m
  \prod (u_i - u_j)^m   \nonumber \\
 & & \hspace{.35in} \times \exp \left[ -\frac{1}{4} \left(
     \sum |z_i|^2 + \sum |u_j|^2 + \sum |w_k|^2 \right) \right] \, .
\label{wvf}
\end{eqnarray}
Here $n$ and $m$ are odd integers, and $l$ can be any integer.
A simple calculation then gives the total filling of this wave function:
$\nu = (n + m - 2 l)/(n m - l^2 )$.
What is remarkable about such a wave function is that, since the
individual filling factors in the $S\downarrow$ and
$A\uparrow$ states are not fixed, we can consider wave functions
that are a superposition of states with
various $N_u - N_w$.  They mix states with different values of
$S^z$ (the $z$ component of spin lies along the direction of the magnetic
field) and therefore describe states with spontaneously broken spin symmetry
--- the CAF state of \cite{R1,R2,R3,R3a,R4,R5,R6,R7}.
It is easy to see that taking $l=0$ and $n=m$ in (\ref{wvf}) gives
$\nu = 2 /m $, i.e. the CAF state discussed in \cite{R1,R2}. In the
CAF phase, the electrons in the two layers have the same $z$ component
of spin but opposite $x-y$ components.  The direction of the Neel
order parameter (defined as the difference in the spin expectation
values in the two layers) comes from the spontaneous breaking of the
$S^z$ spin symmetry.  It should be mentioned that Halperin wave
functions for spontaneously broken spin symmetry states may also be
constructed for single layer QH systems, leading to the possibility
(at least in principle) of exotic spin states in a single layer
QH system \cite{spec}.

Properties of the state (\ref{wvf}) are conveniently discussed using
a bosonic Chern-Simons theory \cite{CS,zhang}.  For simplicity we again
assume that the $A\downarrow$ states are empty and
consider only three kinds of electrons:
$\Psi_1$ for $S\uparrow$, $\Psi_2$ for $S\downarrow$,
and $\Psi_3$ for $A\uparrow$.
Eq.~(\ref{wvf}) tells us that
the electron $\Psi_1$ is seen as a vortex of strength $n$ by
other $\Psi_1$ electrons and a vortex of strength $l$ by electrons
$\Psi_{2,3}$; electrons $\Psi_{2,3}$ are seen as vortices of strength
$m$ and $l$ by electrons $\Psi_{2,3}$ and $\Psi_1$, respectively.
We are therefore led to consider the following (bosonic)
Chern-Simons Lagrangian
\breakon
\begin{eqnarray}
{\cal L} &=& \bar{\Psi}_1 ( \partial_0 - i a_0 ) \Psi_1
+ \sum_{a=2,3} \bar{\Psi}_a ( \partial_0 - i \tilde{a}_0 ) \Psi_a
+ \frac{1}{2 m} \left|
   [ -i \vec{\partial} - n {\bf a} - l \tilde{{\bf a}}
     - {\bf A}^{ex} ] \Psi_1 \right|^2
+ \frac{1}{2 m} \sum_{a=2,3} \left|
   [ -i \vec{\partial} - l {\bf a} - m \tilde{{\bf a}}
     - {\bf A}^{ex} ] \Psi_a \right|^2
\nonumber \\
& & \hspace{.4in} - (\Delta_{Z} + \Delta_{SAS}) \bar{\Psi}_1 \Psi_1
 - ( \Delta_{SAS} -   \Delta_{Z} ) \bar{\Psi}_2 \Psi_2
 + ( \Delta_{SAS} -   \Delta_{Z} ) \bar{\Psi}_3 \Psi_3
\nonumber \\
& & \hspace{1.45in} +u^c_{LL'}(x-y) (\rho_L(x) - \bar{\rho})
             (\rho_{L'}(y) - \bar{\rho} )
  + {\cal L}_{CS}(a) + {\cal L}_{CS}(\tilde{a}) \, ,
\label{L0}
\end{eqnarray}
where $ {\cal L}_{CS} (a) = \frac{i}{ 4 \pi} \epsilon^{\mu \nu
\lambda} a_{\mu} \partial_{\nu} a_{\lambda} $, and $L$ is a layer index
(``Top'' or ``Bottom'') in the Coulomb interaction term.  We
decompose the $\Psi_i$'s into an amplitude, a trivial phase,
and a vortex part \cite{zhang,LeeKane}:
 $ \Psi_1 = \sqrt{ \rho_1} e^{ i \theta_1 } \phi_{v1} $ and
$ \Psi_a = \sqrt{ \rho_2} e^{ i \theta_2 } \phi_{v2} z_{a-1}$ for $a
= 2,~3$, with the constraints $ \bar{\phi}_{v1} \phi_{v1} = \bar{\phi}_{v2}
\phi_{v2} = \bar{z}_a z_a =1$.  Then (\ref{L0}) can be
written as
\begin{eqnarray}
{\cal L} &=& i \rho_1 \left( \frac{ \partial_0 \theta_1}{i}
+  \bar{\phi}_{v1} \frac{ \partial_0 }{i} \phi_{v1} - a_0 \right)
+  i \rho_2 \left( \frac{ \partial_0 \theta_2}{i}
+  \bar{\phi}_{v2} \frac{ \partial_0 }{i} \phi_{v2}
+  \bar{z}_{a} \frac{ \partial_0 }{i} z_{a} - \tilde{a}_0 \right)
\nonumber\\
&+&  i {\bf J} \left( \frac{ \vec{\partial} \theta_1}{i}
+  \bar{\phi}_{v1} \frac{ \vec{\partial} }{i} \phi_{v1} - n {\bf a }
      - l \tilde{{\bf a}} - {\bf A}^{ex}  \right)
+ i \tilde{{\bf J}} \left( \frac{ \vec{\partial} \theta_2}{i}
+  \bar{\phi}_{v2} \frac{ \vec{\partial} }{i} \phi_{v2}
+  \bar{z}_{a} \frac{ \vec{\partial} }{i} z_{a}
  - l {\bf a } - m \tilde{{\bf a}}  - {\bf A}^{ex}  \right)
\nonumber\\
&+& \frac{K_1}{2} \left| {\bf J} \right|^2
    + \frac{K_2}{2} \left| \tilde{{\bf J}} \right|^2
    + \frac{1}{2 K_2} \left( | \vec{\partial} z |^2
    + ( \bar{z}  \vec{\partial} z )^2  \right)
    - (\Delta_{Z} + \Delta_{SAS}) \rho_1
    - ( \Delta_{SAS} - \Delta_{Z} ) \rho_2 | z_1|^2
    + ( \Delta_{SAS} - \Delta_{Z} ) \rho_2 | z_2|^2
\nonumber \\
&-& \sum_{ab} \gamma_{ab} | z_a |^2 | z_b |^2
+ ( \rho_1 + \rho_2 - \bar{\rho} )(x) u(x-y)
       (\rho_1 + \rho_2 - \bar{\rho} )(y)
 + {\cal L}_{CS}(a) + {\cal L}_{CS}(\tilde{a}) \, .
\label{L1}
\end{eqnarray}
Here $K_i = m/ \rho_i$, terms with $\gamma_{ab}$ come from the
exchange part of the Coulomb interaction, and in the direct part
of the Coulomb interaction we keep only the layer symmetric part
of $u^c_{LL'}$ which does not vanish in the limit of $d=0$.
By integrating out $\theta_1$ and $\theta_2$ we find
that $J_{\mu} = ( \rho_1, {\bf J}) $ and $\tilde{J}_{\mu} = ( \rho_2,
\tilde{{\bf J}}) $ are conserved.  Therefore, we introduce dual gauge
fields, $b_{\lambda}$ and $\tilde{b}_{\lambda}$, such that
$
J_{\mu} = \frac{1}{2 \pi} \epsilon^{\mu \nu \lambda} \partial_{\nu} b_{\lambda}
$ and
$
\tilde{J}_{\mu} = \frac{1}{2 \pi} \epsilon^{\mu \nu \lambda}
                  \partial_{\nu} \tilde{b}_{\lambda}
$.
Then, we integrate out the statistical gauge fields, $a_{\mu}$ and
$\tilde{a}_{\mu}$, and the time component of the dual gauge fields,
$b_0$ and $\tilde{b}_0$. This gives (up to irrelevant constants)
\begin{eqnarray}
{\cal L} &=& i b_{\alpha} J^v_{\alpha} + i \tilde{b}_{\alpha} \left(
\tilde{J}^v_{\alpha} +  \tilde{J}^S_{\alpha} \right)
+ \frac{1}{ 8 \pi^2 K_1} \left( \partial_0 b_{\alpha} \right)^2
+ \frac{1}{ 8 \pi^2 K_2} \left( \partial_0 \tilde{b}_{\alpha} \right)^2
+ \frac{1}{2 K_2} \left( | \vec{\partial} z |^2
                     + (\bar{z}  \vec{\partial} z)^2  \right)
+ i \epsilon^{\alpha\beta} b_{\alpha} \partial_0 b_{\beta}
+ i \epsilon^{\alpha\beta} \tilde{b}_{\alpha} \partial_0 \tilde{b}_{\beta}
\nonumber\\
&+& \frac{K_1}{2} r(x) \ln|x-y| r(y)
    + \frac{K_2}{2} \tilde{r}(x) \ln|x-y| \tilde{r}(y)
+ ( r_1 |z_1|^2 + r_2 |z_2|^2 - \gamma_{11} |z_1|^4
    - \gamma_{22} |z_2|^4 - 2 \gamma_{12} |z_1|^2 |z_2|^2 )
\nonumber\\
&+& \frac{1}{4 \pi^2} ( \epsilon^{\alpha \beta} \partial_{\alpha} b_{\beta}
  + \epsilon^{\alpha \beta} \partial_{\alpha} \tilde{b}_{\beta}
  - \bar{\rho} )(x) u(x-y)
    ( \epsilon^{\alpha \beta} \partial_{\alpha} b_{\beta}
  + \epsilon^{\alpha \beta} \partial_{\alpha} \tilde{b}_{\beta}
  - \bar{\rho} )(y) \, ,
\label{L2}
\end{eqnarray}
\breakoff
\noindent
where we have defined vortex and skyrmion currents as in \cite{LeeKane}:
$
J^v_{\mu} = (J^v_0,J^v_{\alpha}) = \frac{1}{2 \pi}
\epsilon^{\mu \nu \lambda} \partial_{\nu} \left( \bar{\phi}_{v1}
\frac{\partial_{\lambda}}{i}
\phi_{v1} \right)
$,
$
\tilde{J}^v_{\mu} = (\tilde{J}^v_0,\tilde{J}^v_{\alpha}) = \frac{1}{2 \pi}
  \epsilon^{\mu \nu \lambda} \partial_{\nu} \left( \bar{\phi}_{v2}
  \frac{\partial_{\lambda}}{i} \phi_{v2} \right)
$
, and
$
\tilde{J}^S_{\mu} = (\tilde{J}^S_0,\tilde{J}^S_{\alpha}) = \frac{1}{2 \pi}
  \epsilon^{\mu \nu \lambda} \partial_{\nu} \left( \bar{z}_{z}
  \frac{\partial_{\lambda}}{i} z_{a} \right)
$.
The parameters $r_1 = - r_2 =  (\Delta_{Z} - \Delta_{SAS}) \rho_2 $,
and
\begin{eqnarray}
r(x) &=& 2 \pi J_0^V - n \epsilon^{\alpha\beta} \partial_{\alpha} b_{\beta}
 - l \epsilon^{\alpha\beta} \partial_{\alpha} \tilde{b}_{\beta}
 - \epsilon^{\alpha\beta} \partial_{\alpha} A^{ex}_{\beta}  \nonumber\\
\tilde{r}(x) &=& 2 \pi \tilde{J}_0^V +
2 \pi \tilde{J}_0^S - l \epsilon^{\alpha\beta} \partial_{\alpha} b_{\beta}
 \nonumber \\
 & & \hspace{.7in}
 - m \epsilon^{\alpha\beta} \partial_{\alpha} \tilde{b}_{\beta}
 - \epsilon^{\alpha\beta} \partial_{\alpha} A^{ex}_{\beta}  \, .
 \nonumber 
\end{eqnarray}

In the ground state there are no vortices or skyrmions, so the
cancellation of the long range logarithmic interaction gives two
conditions
\begin{eqnarray}
\frac{1}{2 \pi} \int d^2 x \left[
\epsilon^{\alpha\beta} \partial_{\alpha} A^{ex}_{\beta}
 + n \epsilon^{\alpha\beta} \partial_{\alpha} b_{\beta}
 + l \epsilon^{\alpha\beta} \partial_{\alpha} \tilde{b}_{\beta} \right]
 = 0 \, , \nonumber\\
\frac{1}{2 \pi} \int d^2 x \left[
\epsilon^{\alpha\beta} \partial_{\alpha} A^{ex}_{\beta}
 + l \epsilon^{\alpha\beta} \partial_{\alpha} b_{\beta}
 + m \epsilon^{\alpha\beta} \partial_{\alpha} \tilde{b}_{\beta} \right]
 = 0 \, .
\label{req}
\end{eqnarray}
Recalling that $1/2 \pi \epsilon^{\alpha\beta} \partial_{\alpha} b_{\beta}$
gives the density of $\Psi_1$ electrons and
$1/2 \pi \epsilon^{\alpha\beta} \partial_{\alpha} \tilde{b}_{\beta} $
gives the density of $\Psi_2$ and $\Psi_3$ electrons, we realize that
Eq.~(\ref{req}) gives us the same filling fractions as the Halperin
wave function (\ref{wvf}).

From the last line of (\ref{L2}) it is obvious that as we
change the strength of the Zeeman interaction and/or interlayer tunneling,
we will stabilize various values of $|z_1|$ and $|z_2|$. The most
important observation is that when $d \neq 0$ we have $ \gamma_{12} >
\gamma_{11}+ \gamma_{22} $, so there is never a direct transition from
$|z_1|=1$ \& $|z_2|=0$ to $|z_1|=0$ \& $|z_2|=1$, but there is always an
intermediate phase where both $|z_1| = \cos \theta_0$ and
$|z_2| = \sin \theta_0$ are finite.  This
corresponds to the CAF phase for fractional fillings. In this phase
interactions fix the absolute values of $z$'s but not their relative
phase.  Therefore, when $\bar{z}_1 z_2$ develops an expectation value,
we have a spontaneous breaking of the $U(1)$ symmetry and the appearance
of a Goldstone mode.

In the CAF phase, dynamics of the spin is determined by
\begin{eqnarray}
 {\cal L}_z & = & i \bar{\rho}_2 \bar{z} \partial_t z - \frac{1}{2 K_2}
 \left( | \vec{\partial} z |^2 + (\bar{z} \vec{\partial} z)^2
 \right)+ r_1 |z_1|^2 + r_2 |z_2|^2 \nonumber \\
 & - & \gamma_{11} |z_1|^4 - \gamma_{22} |z_2|^4
   - 2 \gamma_{12} |z_1|^2 |z_2|^2  \, .
\label{Lz}
\end{eqnarray}
In the CAF phase only
$\langle \bar{z}_1 z_2 \rangle $  develops a non-zero expectation value,
but not $\langle z_1 \rangle $
or $\langle z_2 \rangle $. Therefore, we can write
$ z_1 = | z_1 | e^{ i ( \phi + \chi ) }$ and $ z_2 = |
z_2 | e^{ i ( - \phi + \chi ) }$.  $\phi$, the relative phase between
$z_1$ and $z_2$, acquires an expectation value and gives rise to
the Goldstone mode associated with the symmetry breaking.
We also introduce $q = | z_1 |^2 - | z_2 |^2$,
where far away from a vortex core $ q_{{\rm min}} = \cos
( 2 \theta_0)$. Using that fluctuations of $q$
are massive and their gradients may be neglected, we find from (\ref{Lz})
\begin{eqnarray}
 {\cal L}_z & = & i \bar{\rho}_2 \delta q~ \partial_t \phi
              - \frac{1}{2 K_2} ( \vec{\partial} \phi )^2
          [ 1 - q_{{\rm min}}^2 ]   \nonumber \\ 
        & + & \alpha ( \delta q )^2 
          + i \bar{\rho}_2~ q_{{\rm min}} \partial_t \phi \, ,
\end{eqnarray}
with $\alpha = ( \gamma_{11} + \gamma_{22} - 2 \gamma_{12} )/4  $.
Integrating out $\delta q$ we get
\begin{eqnarray}
{\cal L}_{\phi} & = & \frac{\bar{\rho}_2^2}{4 \alpha^2}
             \left( \partial_t \phi \right)^2
- \frac{1}{2 K_2} ( \vec{\partial} \phi )^2 [ 1 - q_{{\rm min}}^2 ]  
\nonumber \\
 &  & \hspace{.2in} + i \bar{\rho}_2~ q_{{\rm min}}  \partial_t \phi \, .
\end{eqnarray}
We see that the spin wave velocity is
$v_s =  ( \gamma_{11} + \gamma_{22} - 2 \gamma_{12} )^2
        \sin^2 ( 2 \theta_0 ) / (8 m \bar{\rho}_2)$.
By introducing an infinitesimal external Zeeman field
and integrating out fluctuations in $\phi$, one
can also calculate the $S^z$ correlation function
which explicitly shows a Goldstone resonance
\begin{eqnarray}
\chi^{zz}(q,\omega) =  \frac{
  \frac{ \bar{\rho}_2^2}{2\alpha} \omega^2  }
 { \omega^2 - v_s^2 k^2 }
  - \frac{ \bar{\rho}_2^2}{2\alpha}  \, .
\end{eqnarray}

When the ground state of a system breaks a $U(1)$ symmetry spontaneously,
vortices of the $U(1)$ phase will be the elementary excitations in the
system. For the $(1,1,1)$ states discussed in \cite{WenZee,Indiana1},
these elementary excitation are merons which were shown to have fractional
charge \cite{Indiana1}.  However, a pair of merons with opposite vorticity
always add up to integer charge, $0$ or $1$. For the wave functions defined
in (\ref{wvf}), one can also imagine a meron in which the direction of the
Neel vector points outward from the center of the vortex, and ask whether
such a meron will carry an electric charge \cite{meron}.
It is clear from the discussion above that such a meron corresponds to a
vortex of the $z$ field. Far away from the vortex core, $|z_1| = \cos
\theta_0 $ and $|z_2| = \sin \theta_0 $, and the relative phase between
the two $z's$ has nontrivial winding characterized by integer
vorticity, $n_v$. In order to avoid a singularity of this phase,
in the vortex core we must have either $ |z_1| = 1$ \& $
|z_2| = 0$ (S vortex) or $ |z_1| = 0$ \& $ |z_2| =
1$ (T vortex). According to the definition of
$\tilde{J}^S_0$, this implies a nontrivial skyrmion winding number
\begin{eqnarray}
Q^s & = & \frac{1}{2 \pi} \int d^2 x \tilde{J}^{S}_0
    = \frac{1}{2 \pi} \int d^2 x \epsilon^{\alpha \beta}
     \partial_{\alpha} \left( \bar{z} \frac{\partial_{\beta}}{i} z \right)
 \nonumber \\
 & = & \left\{ \begin{array}{c} \ \, n_v \times
                           \sin^2 \theta_0~~~~~S~ {\rm vortex}
\\ - n_v \times \cos^2 \theta_0 ~~~~T~ {\rm vortex} \end{array} \right. \, ,
\label{Qseq}
\end{eqnarray}
where $n_v$ is an integer characterizing the winding of the relative phase
between $z_1$ and $z_2$.
From (\ref{L2}) the extra skyrmion charge has to be compensated by
electric charge.  To cancel the long-range forces,
$  \int [ -n \epsilon^{\alpha \beta} \partial_{\alpha} b_{\beta}
 -l \epsilon^{\alpha \beta} \partial_{\alpha} \tilde{b}_{\beta} ] = 0
$
and
$
 \int [ -l \epsilon^{\alpha \beta}
         \partial_{\alpha} b_{\beta} -m \epsilon^{\alpha \beta}
         \partial_{\alpha} \tilde{b}_{\beta} ] =  - 2 \pi Q^s$,
which immediately gives us the total charge of the meron
\begin{eqnarray}
Q_{{\rm meron}} & = & \frac{1}{2 \pi} \int \left[ \epsilon^{\alpha \beta}
   \partial_{\alpha} b_{\beta}
+ \epsilon^{\alpha \beta} \partial_{\alpha} \tilde{b}_{\beta} \right]
 \nonumber \\
   & = & \frac{ n- l}{n m - l^2} \times Q^s \, .
\label{Qmeron}
\end{eqnarray}
For $l=0$, which includes the $\nu = 2/m$ states discussed in
\cite{R2}, we find $Q_{{\rm meron}} = 1/m \times Q^s$.
Note that if we were
to create simple quasiparticles by squezing a vortex into the ground
state, $J^v_0 = \delta^2 ( x - x_0 )$ or
$\tilde{J}^v_0 = \delta^2 ( x - x_0 )$, we could use the same arguments
to find their charges: $ q = (m - l)/(n m - l^2)$ and
$ \tilde{q} = (n - l)(n m - l^2) $.
So, as in the simple case of a meron in the $(m,m,m)$ state,
two merons add up to a charge of $0$ or the charge of
a single quasiparticle, $\tilde{q}$ \cite{eduardo}.

In the simplest case of $n=m=1$ and $l=0$ in (\ref{wvf}) (i.e. $\nu=2$),
one can give a simple picture of the meron excitation in the CAF phase
using a generalization of the Berry's phase argument in
\cite{Indiana1}.  As suggested in \cite{R3}, the CAF phase can be
described by combining pairs of electrons into hard core bosons and
writing the wave function as $ | \Psi \rangle = \cos \theta | S \rangle
+ e^{i \phi} \sin \theta | T \rangle $.  Here $|S\rangle$ and 
$|T\rangle$ denote singlet and triplet bosons, respectively; the relative 
phase between the two bosons, $\phi$, determines the direction of the 
Neel vector in the $x-y$ plane. When a vortex is present, this phase winds
nontrivially around the vortex core and is characterized by an integer
vorticity, $n_v$.  At the center of the core, one has to demand that
there is only one kind of boson present (so as to avoid a singularity
of the relative phase); therefore, one expects the appearance of
two kinds of vortices: vortices with a singlet core (S) or a triplet
core (T).  We can now imagine taking a pair of electrons and
adiabatically moving them around the vortex. In the course of such
adiabatic transport, the wave function for a pair of electrons will
acquire a phase
$i \Gamma = \oint \langle \psi | d \psi \rangle = i~n_v \sin^2 \theta$
for an S vortex or $i \Gamma = - i~n_v \cos^2 \theta$ for a T vortex.
The Berry's phase in adiabatic transport is indistinguishable from extra
flux going through the system $ \Delta \Phi = \Phi_0/2 \times \Gamma/
(2 \pi) $, where the factor of $1/2$ comes from the fact that we
transported a pair of electrons.  This extra flux can be related
to the charge carried by the meron as $ \Delta q = \sigma_{xy} \Delta
\Phi =  n_v \times \sin^2 \theta $ and
$ \Delta q = - n_v \times \cos^2 \theta $ for $S$ and $T$ vortices,
respectively. So, the two kinds of merons in this case carry
fractional charge; the charge depends on where the system is in the
phase diagram, i.e. on the CAF phase order parameter $\theta$ (
$\theta$ goes to $0$ at the boundary of the CAF phase with the spin
singlet phase, and $\pi/2$ at the boundary with the fully polarized
ferromagnetic state; see \cite{R3} for details ). However, two
merons with opposite vorticities again add up to a charge of $0$ or $1$,
as in the $\nu=1$ bilayer $(1,1,1)$ state \cite{Indiana1}.

It is also instructive to consider an explicit wave function for a
meron in the CAF phase at $\nu=2$. As discussed above, the wave
function of the CAF phase may be conveniently written as (in the
limit when $d$ is small)
$
| \Psi_0 \rangle = \prod_m ( \cos \theta~S^{\dagger}_m
   + e^{i\phi} \sin \theta~T^{\dagger}_m )  | 0 \rangle
\label{Psi0}
$,
where $ S^{\dagger}_m = 1/\sqrt{2} ( c_{Sm\uparrow}^{\dagger}
 c_{Sm\downarrow}^{\dagger} - c_{Sm\downarrow}^{\dagger}
 c_{Sm\uparrow}^{\dagger} )$ and $T_m^{\dagger} =
 c_{Sm\uparrow}^{\dagger} c_{Am\uparrow}^{\dagger} $ create singlet
 and triplet combinations of electrons with orbital momentum
 $m$, and $| 0 \rangle$ is the Fock vacuum.
Using the definition of the Neel order parameter,
$
N^a(z) = \langle \Psi | S_T^a - S^a_B | \Psi \rangle
  = \sum_{mn} \Psi^{*}_m(z) \Psi_n (z)
  \times \langle \Psi | c_{Sm\alpha}^{\dagger}
  \sigma^a_{\alpha\beta} c_{An\beta} - c_{Am\alpha}^{\dagger}
  \sigma^a_{\alpha\beta} c_{Sn\beta} | \Psi \rangle \, ,
$
where $\Psi_m(z)$ is the wavefunction of an electron in the first 
Landau level with angular momentum $m$, one can easily prove that 
state $|\Psi_0\rangle$ has a uniform ${\vec N}$ in the $XY$ plane
$
N^{+}(z)  =  1/2 \cos \theta \sin \theta e^{ i \phi}
$.
To have a meron we need a wave function where the
direction of the Neel vector winds around as one goes around the
center of the meron. This is achieved by considering the following
wave function
$
| \Psi_M \rangle =  \prod_m ( \cos \theta~S^{\dagger}_m
  + e^{i\phi} \sin \theta~\tilde{T}^{\dagger}_m )  | 0 \rangle
\label{PsiM}
$,
where $\tilde{T}_m^{\dagger} =
c_{Sm+1\uparrow}^{\dagger} c_{Am\uparrow}^{\dagger} $.
For $|\Psi_M\rangle$
one finds that $N^{+}(z) = 1/2 \sum_m \Psi^{*}_m (z) \Psi_{m+1} (z)
\cos \theta \sin \theta e^{ i \phi_0}$.  Since $\Psi_m \propto z^m \exp( -
|z|^2)$ we find that
$
\arctan(N_y/N_x) = \arg(z) + \phi
$;
the direction of $\vec{N}$ winds in the $XY$ plane following the
argument of the complex coordinate $z$. It is also obvious from
$| \Psi_M \rangle$ that it describes a state with a missing
electron in the $S\uparrow$ state of $ m=0$, so we have an $S$ vortex
with charge $ - \sin^2 \theta$.

In summary, we have developed an analytic theory for the bilayer QH
CAF phase.  Our theory is consistent with the original Hartree-Fock theory
for $\nu=2$, but is general enough to predict a whole new class of {\it
fractional} QH CAF phases as well as the correct excitation spectra.

We thank M.P.A. Fisher, A.H. MacDonald, T. Senthil and
K. Yang  for useful discussions, and B.I. Halperin and 
S.M. Girvin for a careful reading of the manuscript.
This work was supported by the National Science Foundation (ED), the
US Department of Energy under Grant No. DE-FG03-85ER45197 (EHK),
and ONR (SDS).


\end{multicols}

\begin{thebibliography}{10}

\vspace{-.7in}

\bibitem{R1}
L. Zheng {\it et. al. }, {\em Phys. Rev. Lett.}, {\bf 78}:2453 (1997).

\bibitem{R2}
 S. Das Sarma {\it et. al. }, {\em Phys. Rev. Lett.}, {\bf 79}:917 (1997);
{\em Phys. Rev. B} {\bf 58}:4672 (1998).

\bibitem{R3} E. Demler and S. Das Sarma, {\em Phys. Rev. Lett.}, {\bf
82}:3895 (1999).

\bibitem{R3a}   Kun Yang, cond-mat/9903437.

\bibitem{R4} L. Brey, E. Demler, and S. Das Sarma, {\em Phys. Rev. Lett.}, {\bf
83}:168 (1999).

\bibitem{R5} L. Brey, {\em Phys. Rev. Lett.}, {\bf
81}:4692 (1999); B. Paredes {\it et. al.}, cond-mat/9902197.

\bibitem{R6} A.  MacDonald {\it et. al.}, cond-mat/9903318.

\bibitem{R7} V. Pellegrini {\it et. al.}, {\em Phys. Rev. Lett.}, {\bf
79}:310 (1997); {\em Science}, {\bf 281}:799 (1998).

\bibitem{WenZee} X.G. Wen and A. Zee , {\em Phys. Rev. Lett.}, {\bf
69}:1811 (1992).

\bibitem{Halperin} B. I. Halperin, {\em Helv. Phys. Acta },
{\bf 56}:75 (1983).

\bibitem{Indiana1} K. Moon {\it et. al.},  {\em Phys. Rev. B},
{\bf 51}:5138 (1995).

\bibitem{note}
The generalization to include electrons in the $A\downarrow$ state is
straightforward.

\bibitem{spec} S. Das Sarma and E. Demler, unpublished.

\bibitem{CS}
S.C. Zhang, T.H. Hansson, and S. Kivelson,
{\em Phys. Rev. Lett.}, {\bf 62}:82 (1989).

\bibitem{zhang} 
S.C. Zhang {\em Int. J. Mod. Phys. B}, {\bf 6}:25 (1992).

\bibitem{LeeKane}
D.H. Lee and C.L. Kane, {\em Phys. Rev. Lett.}, {\bf 64}:1313 (1990).

\bibitem{meron}
We use the term meron, since a vortex with a triplet core
corresponds to a spin configuration in each layer that is similar 
to a pseudospin meron in $\nu=1$ bilayer QH systems.

\bibitem{eduardo}
A. Lopez and E. Fradkin in {\em Composite Fermions}, edited by
O. Heinonen (World Scientific, Singapore, 1998).


\end{thebibliography}
\end{document}